\documentclass[journal]{IEEEtran}

\usepackage{graphicx}   
\usepackage{cite}       
\usepackage{url}
\usepackage{amsmath}
\usepackage{array}

\usepackage{helvet}  
\usepackage{tikz}
\usetikzlibrary{arrows.meta,positioning,shapes.geometric,calc,shapes.misc,backgrounds,fit}
\definecolor{cRAN}{HTML}{1F6FB2}
\definecolor{cSA}{HTML}{2E8B7F}
\definecolor{cCT}{HTML}{C8772E}
\definecolor{cInk}{HTML}{2B2B2B}
\definecolor{cGrey}{HTML}{7A828C}
\definecolor{cLight}{HTML}{EDF3F8}
\definecolor{cRel20}{HTML}{9CC3DE}
\definecolor{cRel21}{HTML}{1F6FB2}
\definecolor{cITU}{HTML}{B9Bec6}
\definecolor{cBanderL}{HTML}{CFE3F1}
\definecolor{cBanderR}{HTML}{3F86BE}

\usepackage{xurl}
\usepackage{hyperref}
\hypersetup{
     colorlinks   = true,
     citecolor    = red,
     linkcolor    = red,
     urlcolor     = black
}

\usepackage[table]{xcolor}
\definecolor{myblue}{RGB}{217,226,243}

\newcommand{\stag}[1]{\textsuperscript{\sffamily\bfseries\scriptsize #1}}
\newcommand{\lesson}[2]{%
  \par\addvspace{4pt}\noindent
  \fcolorbox{black!75!black}{blue!5}{%
    \begin{minipage}{\dimexpr\columnwidth-2\fboxsep-2\fboxrule\relax}
      \footnotesize\color{black}\textbf{Lesson (#1).}\ #2%
    \end{minipage}}%
  \par\addvspace{4pt}}

\usepackage{balance}

\usepackage{eso-pic}
\begin{document}

\AddToShipoutPictureBG*{%
  \AtPageUpperLeft{%
    \put(\LenToUnit{0.5\paperwidth},\LenToUnit{-1.2cm}){%
      \makebox[0pt]{\parbox{0.9\paperwidth}{\scriptsize\centering
        \copyright~2026 IEEE. Personal use of this material is permitted. Permission from IEEE must be obtained for all other uses, in any current or future media, including reprinting/republishing this material for advertising or promotional purposes, creating new collective works, for resale or redistribution to servers or lists, or reuse of any copyrighted component of this work in other works.}}}}}

\title{6G as It Is Actually Being Built: Insights from Early 3GPP Standardization}

\author{Ngo~Hoang~Tu,~\IEEEmembership{Member,~IEEE},
        Huy~T.~Nguyen,~\IEEEmembership{Member,~IEEE}, 
        Mao~V.~Ngo,~\IEEEmembership{Member,~IEEE},
        Tran~Thien~Thanh,~\IEEEmembership{Member,~IEEE},
        Vo~Nguyen~Quoc~Bao,~\IEEEmembership{Senior Member,~IEEE},
        and
        Tony~Q.~S.~Quek,~\IEEEmembership{Fellow,~IEEE}
\thanks{Ngo Hoang Tu, Huy T. Nguyen, and Vo Nguyen Quoc Bao are with the Faculty of Information Technology, Van Lang School of Technology, Van Lang University, Ho Chi Minh City 70000, Vietnam (e-mail: tu.nh@vlu.edu.vn, huy.nt@vlu.edu.vn, bao.vnq@vlu.edu.vn).}
\thanks{Mao V. Ngo is with the Future Communications Research \& Development Programme, Singapore University of Technology and Design, Singapore 487372, Singapore (e-mail: vanmao\_ngo@sutd.edu.sg).}
\thanks{Tran Thien Thanh is with the Institute of Information Technology and Electrical-Electronics Engineering, Ho Chi Minh City University of Transport, Ho Chi Minh City 70000, Vietnam (e-mail: thanh.tran@ut.edu.vn). (\textit{Corresponding author: Tran Thien Thanh.})}
\thanks{Tony Q. S. Quek is with the Information Systems Technology and Design Pillar, Singapore University of Technology and Design, Singapore 487372, Singapore (e-mail: tonyquek@sutd.edu.sg).}
}


\maketitle

\begin{abstract}
As mobile communications cross the threshold from fifth-generation (5G) to sixth-generation (6G), the 3rd Generation Partnership Project (3GPP) has entered a decisive phase. 
Following the Technical Specification Group (TSG) plenary meetings of June 2026 and the associated 6G workshop in Singapore, the Release-20 study phase is now well underway across all three TSGs: Radio Access Network (RAN), Service and System Aspects (SA), and Core Network and Terminals (CT). Concurrently, the timeline for the first normative 6G specifications in Release 21 has been established. 
Rather than offering another technology vision, or a chronological account of committee progress, this article asks what the first body of agreed 6G material tells us. We review service requirements, system architecture, the emerging radio interface, native Artificial Intelligence (AI), integrated sensing and communication (ISAC), non-terrestrial networks (NTN), and core-network protocols, tracing how each moves from study item toward specification. Read across the three TSGs, this material also supports seven cross-cutting lessons: AI has moved from a feature to a primitive spanning every stage, non-terrestrial access has crossed from overlay to substrate, and connectivity is now one objective among three alongside computing and sensing; the industry favors deployability over redesign, with spectrum access and site reuse, rather than waveform innovation, appearing to be the tighter constraint; and standardization targets AI workflows rather than AI models, with testability across vendors recurring as a decisive criterion for which techniques survive. Because the material spans very different stages of maturity, every item cited is tagged with its status at a single snapshot date (approved, draft, individually proposed, or vendor-demonstrated), and the lessons are offered as the authors' synthesis rather than as 3GPP positions. We close with the road toward trials and commercial launch around 2030, and with the open questions these lessons help prioritize.
\end{abstract}


\IEEEpeerreviewmaketitle

\vspace{-0.25cm}
\section{Introduction}
\IEEEPARstart{E}{very} generation of mobile communication has followed a recognizable rhythm: a vision crystallizes, requirements are agreed, an architecture is drawn, protocols are written, and networks are trialed and commercialized. The transition from fifth-generation (5G) to sixth-generation (6G) systems is no exception, but it is unfolding with unusual visibility. The International Telecommunication Union Radiocommunication Sector (ITU-R) has set the calendar through its International Mobile Telecommunications (IMT)-2030 framework \cite{itur}, targeting technology proposals around early 2029, and the 3rd Generation Partnership Project (3GPP) is developing the candidate technology. 
To meet these dates while continuing to evolve 5G-Advanced, 3GPP has concluded that two releases are needed to specify 6G: Release 20, carrying the 6G study phase alongside continued 5G-Advanced work, and Release 21, delivering the first set of 6G technical specifications and the IMT-2030 submission \cite{lin20}.

Mid-2026 is therefore a critical inflection point. At the June 2026 plenary round and the associated 6G workshop in Singapore, the Radio Access Network (RAN)-level study on 6G scenarios and requirements was approved \cite{tsg112,tr38914}, the schedule for Release-21 normative work was agreed, and the parallel studies in Service and System Aspects (SA) and Core Network and Terminals (CT) reached their first consolidated baselines.
This marks the first moment at which a reasonably complete, cross-Technical Specification Group (TSG) picture of the 6G system can be drawn directly from agreed 3GPP material rather than from speculation.

The 6G literature is already rich in forward-looking visions and technology surveys \cite{wang23,chen2026toward}, and excellent overviews of the Release-19 and Release-20 packages exist \cite{lin20,lin19}. 
This article takes a deliberately different, and deliberately timely, vantage point. Instead of asking what 6G will be, it asks how 6G is being made, tracking the study-to-specification machinery across RAN, SA, and CT and reporting the agreed status of each technical pillar as of the June 2026 milestones. Existing overviews follow a single release package or a single working group; the three workstreams are read here together, at the one point where they share a common baseline and before the Release-21 work items pull them apart. That vantage point also yields insights no single workstream can, since the same design tensions recur in the RAN spectrum debate, the SA2 architecture baseline, the scoping of native AI, and the CT protocol re-examination. 
Where that recurrence supports a general observation, we state it as a lesson beside the agreements it rests on, and Table~\ref{tab:lessons} groups these under what the system is becoming, how it is being built, and what decides adoption. They are offered as interpretation rather than as 3GPP positions, and as one useful reading of the material rather than its only one. Because the material read here spans very different stages of maturity, every item cited in the sections that follow carries a status tag fixed to a single snapshot date; Section~\ref{sec:process} defines that tag and states the criteria by which an observation was raised to a lesson.

\begin{figure*}[!t]
\centering
\fbox{\includegraphics[width=.985\linewidth]{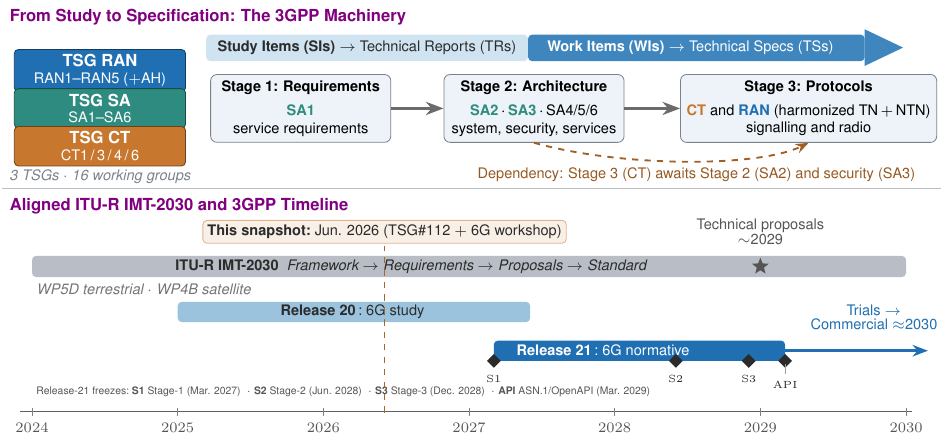}}
\caption{Two views of the 6G standardization program: 
(\textit{i})
The study-to-specification machinery (top panel): SIs and WIs carry TRs and TSs through the three-stage flow (requirements, architecture, protocols) across the RAN, SA, and CT TSGs and their 16 WGs; the principal cross-TSG dependency is that Stage-3 protocol work in CT awaits the Stage-2 architecture (SA2) and security (SA3);
and
(\textit{ii})
The aligned ITU-R IMT-2030 and 3GPP timeline (bottom panel): the Release-20 study phase and the Release-21 normative phase, the four Release-21 freeze milestones (S1, S2, S3, and the interface-definition freeze for Abstract Syntax Notation One and OpenAPI), the June 2026 snapshot on
which this article is based, and the arc toward trials and commercial launch
around 2030.}
\label{fig:overview}
\end{figure*}

\section{From Study Item to Specification: How 3GPP Builds 6G}
\label{sec:process}
3GPP is a consensus-driven partnership in which companies, not individuals, participate on equal terms, and decisions are taken only when there is ``no sustained objection.'' Work is organized into Study Items, which assess feasibility in Technical Reports (TRs), and Work Items, which produce Technical Specifications (TSs). Features mature through three overlapping stages: Stage~1 captures service requirements, Stage~2 defines the system architecture, and Stage~3 specifies protocols and signaling. Sixteen Working Groups (WGs) execute this work and submit it to their parent TSG. The top panel of Fig.~\ref{fig:overview} summarizes this machinery.

\textit{Evidence base and status labels.} The material surveyed here sits at very different stages of maturity, and reporting it uniformly would overstate what has been settled. Every item cited in Sections~\ref{sec:req}--\ref{sec:ct} therefore carries a tag fixed to one snapshot date, 30 June 2026: \stag{A} approved by the parent group and under change control; \stag{D} draft output, whether a study report or normative text, agreed at working-group level but not yet approved; \stag{P} an individual contribution or work-item description, submitted but not agreed; and \stag{T} a vendor or operator trial, cited as feasibility evidence rather than as a 3GPP result. Dates quoted for \stag{D} items are agreed targets, not commitments, and \stag{P} and \stag{T} items never stand as evidence of consensus. The same key is repeated at the foot of Fig.~\ref{fig:ai} and Table~\ref{tab:lessons}.

Release 20 is deliberately dual-track: roughly half of the available capacity advances 5G-Advanced normative work, the other half launches the 6G studies. The 6G radio work in RAN proceeds in three steps: 
\begin{itemize}
    \item A RAN-level study on scenarios and requirements (TR~38.914\stag{A}), approved at TSG RAN\#112, 8--12 June 2026 \cite{tr38914,tsg112};
    \item A WG-level study on the 6G radio interface (initiated in June 2025 and due in June 2027);
    \item The normative Work Item phase, scheduled to run from March 2027 to December 2028.
\end{itemize}
Release 21 will be delivered as a single drop with a unified code freeze, the definitive vehicle for the IMT-2030 submission before 2030 \cite{lin20}. Its agreed schedule \cite{tsg112} places the Stage-1 freeze in March 2027, Stage-2 in June 2028, Stage-3 in December 2028, and the interface-definition freeze in March 2029. The bottom panel of Fig.~\ref{fig:overview} aligns these phases with the ITU-R IMT-2030 schedule and marks the June 2026 snapshot on which this article is based.

A defining feature of this dual-release transition is cross-layer dependency. 
Protocol work in CT cannot be finalized until the Stage-2 architecture in SA2 and the security design in SA3 are agreed; the RAN radio design must stay harmonized with terrestrial and non-terrestrial deployment constraints from the outset; and virtually every WG must coordinate on how native Artificial Intelligence/Machine Learning (AI/ML) integrates into its domain. The 6G program is therefore best read as a coordinated funnel of parallel studies converging on a single, interoperable specification.

\textit{How the lessons were derived.} The seven lessons are our synthesis rather than 3GPP conclusions, and follow from a simple procedure. Each technical pillar was read against three questions: what is treated as foundational rather than optional, what is carried over from 5G rather than redesigned, and which criterion is invoked when scope has to be cut. A recurring answer became a lesson only if it appeared in material from at least two of the three TSGs, or in both a requirements-stage and a radio- or protocol-stage item, \emph{and} rested on at least one \stag{A} or \stag{D} output rather than on \stag{P} or \stag{T} evidence alone. Patterns visible in a single workstream are left in that section's text, and Table~\ref{tab:lessons} lists the qualifying evidence, with its status, behind each lesson.

\begin{figure*}[!t]
\centering
\fbox{\includegraphics[width=.985\linewidth]{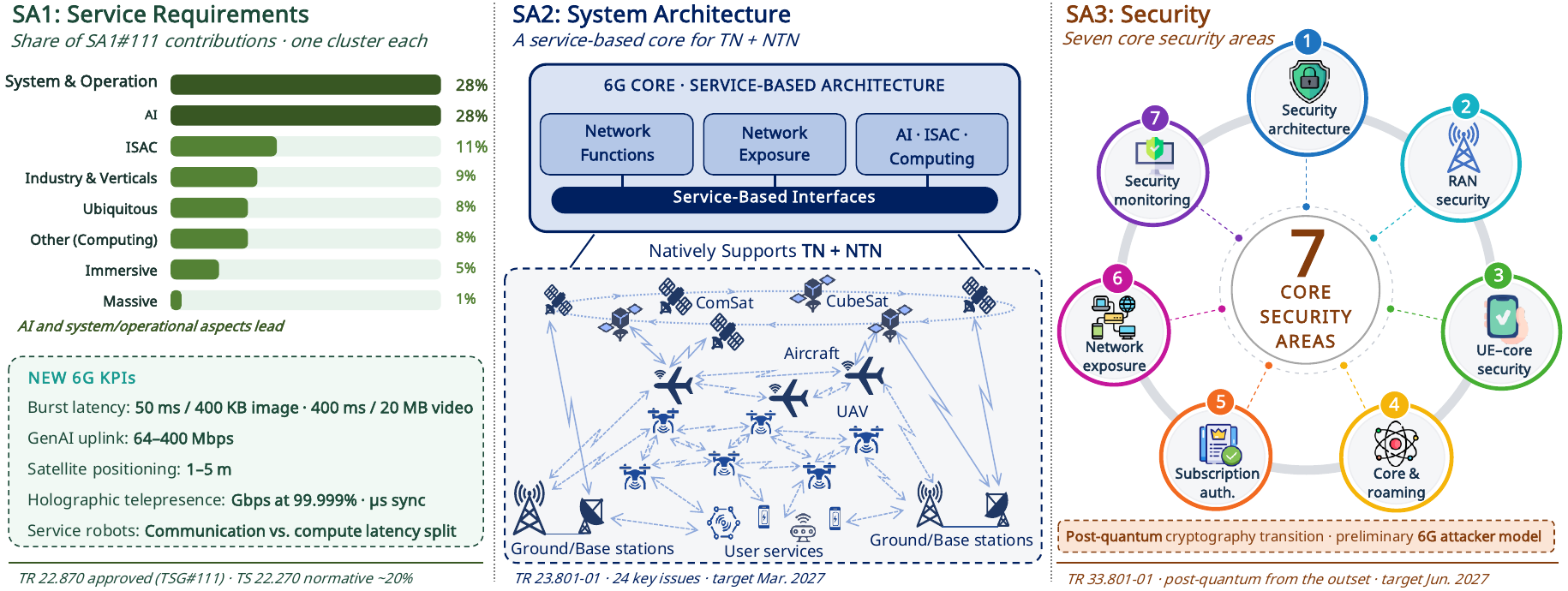}}
\caption{Defining 6G across the SA groups:
(\textit{i}) The SA1 service-requirements study (TR~22.870\stag{A}): contributions by requirement cluster, with system and operational aspects and AI being the most active, together with representative new KPIs;
(\textit{ii}) The SA2 system-architecture study (TR~23.801-01\stag{D}): a service-based core that natively supports terrestrial networks (TNs) and non-terrestrial networks (NTNs);
and
(\textit{iii}) The SA3 security study (TR~33.801-01\stag{D}): seven core security areas, developed with a post-quantum transition from the outset.
\textit{Counting method for the left panel.} The denominator is the full set of contribution documents treated under the 6G service-requirements agenda item at SA1\#111, as tabulated in the SA1 status report presented at the June 2026 workshop \cite{keysight_workshop}. Each document is assigned to exactly one of the eight requirement clusters used by the study itself, so no document is counted twice and the eight shares partition one denominator; shares are rounded and need not sum to 100. The panel therefore measures the relative volume of company interest, not agreed content: no share is a 3GPP position, and it is the requirement text approved in TR~22.870\stag{A} that carries normative weight.}
\label{fig:req}
\end{figure*}

\section{Defining 6G: Service Requirements and System Architecture}
\label{sec:req}
The Stage-1 foundation represents the most mature phase of the current 6G standardization program. The SA1 study on 6G use cases and service requirements (TR~22.870\stag{A}) was completed and approved at the TSG\#111 plenary in March 2026, and is the one SA 6G study output already under change control. By June 2026, the normative specification TS~22.270\stag{D} had passed 20\% completion, with 163 consolidated potential requirements in the baseline text.
The requirements span eight clusters: system and operational aspects, AI, integrated sensing and communication (ISAC), ubiquitous connectivity, immersive communication, massive communication, industry and verticals, and compute-centric networking. System operations and native AI have attracted the highest volume of company contributions.

What is striking is how concrete the new key performance indicators (KPIs) have become. Rather than resting on abstract orders of magnitude, SA1 specifies burst-latency targets tailored to generative-AI demands (50~ms for a 400~KB image-generation burst, 400~ms for a 20~MB video burst), uplink throughput of 64--400~Mbps for edge-rendered AI, and a service-latency breakdown separating communication from compute latency for collaborative robots. Ubiquitous-connectivity requirements add satellite positioning of 1--5~m and satellite-navigation-independent timing, while immersive cases reach toward gigabit holographic telepresence at 99.999\% reliability. AI, sensing, and non-terrestrial access are treated as first-class architectural citizens, not late-stage add-ons.

On the architecture side, the SA2 study (TR~23.801-01\stag{D}) has identified 24 key issues spanning control-plane signaling, quality of service, and native support for sensing and distributed computing. Taking an evolved Service-Based Architecture (SBA) as its baseline, it assumes that the 6G core integrates terrestrial and non-terrestrial networks from day one, with completion targeted for March 2027. Security is developed in parallel by SA3 (TR~33.801-01\stag{D}) across seven domains, from security architecture and RAN security to subscription authentication and security monitoring, including a preliminary attacker model and an explicit post-quantum roadmap. The remaining groups complete the ecosystem: media streaming in SA4 (TR~26.870\stag{D}), intent-driven management in SA5 (TR~32.801-01\stag{D}), and application enablement in SA6 (TR~23.801-02\stag{D}). Of the SA outputs, only TR~22.870 is approved; the rest are draft baselines whose key issues can still be added to, merged, or dropped. Fig.~\ref{fig:req} pairs the SA1 clusters and headline KPIs (left) with the SA2 SBA baseline (center) and the SA3 security domains (right).

\lesson{Connectivity is now one objective among three}{Compute-centric networking is a Stage-1 requirement cluster in its own right, SA1 splits robot service latency into a communication part and a compute part, and SA2 lists native sensing and distributed computing among its key issues. Results stated purely in bits per second will increasingly be read against joint communication, computing, and sensing objectives. Scope: this reflects how the requirements are framed, not an agreed 3GPP resource model.}

\begin{figure*}[!t]
\centering
\fbox{\includegraphics[width=.98\linewidth]{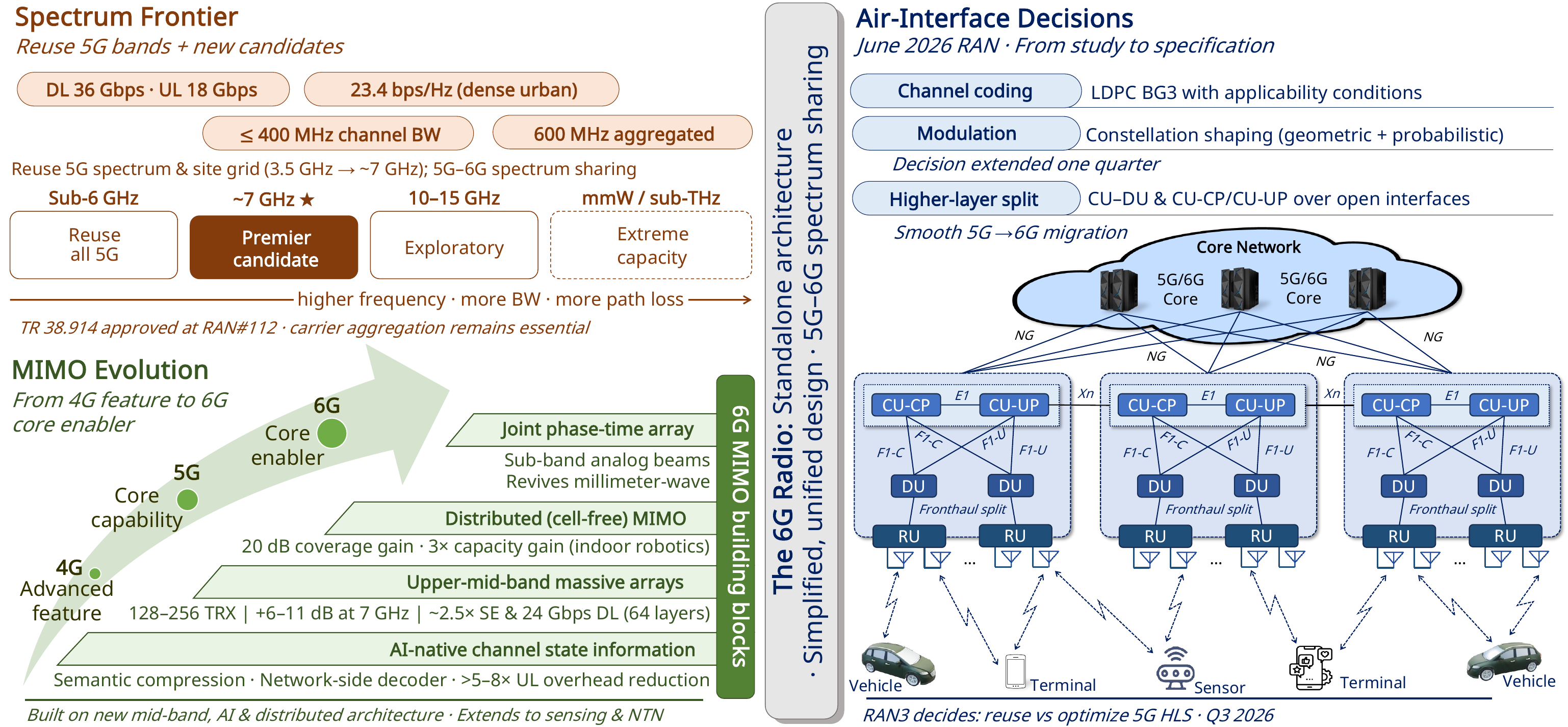}}
\caption{The 6G radio in one view: the spectrum frontier; the June-2026 RAN decisions; and the MIMO evolution from fourth-generation (4G) to 6G.
Of the three air-interface items in the right-hand panel, only the channel-coding decision is an agreed 3GPP outcome\stag{A}: constellation shaping was deferred to September 2026\stag{P}, and the higher-layer-split choice was delegated to RAN3 for the third quarter of 2026. The performance figures in the upper-left panel are agreed\stag{A}. In the MIMO panel, the 2.5-fold spectral efficiency (SE), 24~Gbps at 64 layers, 20~dB coverage gain, and 3-fold capacity gain are vendor field trials\stag{T} \cite{samsungfr3,samsung7ghz,onggosanusi-mimo} and the more than five-fold feedback reduction is the individual contribution R1-2605065\stag{P}; all are included as feasibility evidence and not as 3GPP consensus. TRX denotes transceivers.}
\label{fig:radio}
\end{figure*}

\section{The 6G Radio Interface}
\label{sec:radio}
The RAN-level study on 6G scenarios and requirements (TR~38.914\stag{A}) was approved at TSG RAN\#112 in June 2026 \cite{tr38914,tsg112}, and now provides the industry's first agreed performance baselines for the 6G radio interface. These include a downlink (DL) peak data rate of 36~Gbps and an uplink (UL) rate of 18~Gbps over 600~MHz of aggregated bandwidth, a DL spectral efficiency target of 23.4~bps/Hz in dense-urban environments, and support for individual channel bandwidths up to 400~MHz on the DL. 
Crucially, a core design goal is the strict reuse of the existing 5G site grid, deploying 6G at around 7~GHz to match the coverage of 5G at 3.5~GHz, with a standalone architecture whose simplified, unified frame design natively enables dynamic 5G-6G spectrum sharing.

At the June 2026 RAN plenary meeting, several definitive decisions were made that illustrate this structural shift from study to specification. 
For channel coding, RAN agreed\stag{A} to support a low-density parity-check base graph 3 (LDPC BG3) subject to applicability conditions, while the evaluation of geometric and probabilistic constellation shaping was deferred to the September 2026 plenary and so remains open\stag{P}. On the higher-layer split (HLS), RAN affirmed\stag{A} that 6G must support a split ensuring smooth migration from 5G, covering central-unit/distributed-unit/radio-unit and control-plane/user-plane separation over standardized open interfaces, and tasked RAN3 with deciding by the third quarter of 2026 whether to inherit the 5G split or optimize it. RAN1 was concurrently asked for an interim assessment of waveform, modulation, and coding enhancements beyond the new radio (NR) baseline, together with channel bandwidth, numerology, and synchronization-signal structure. The migration options themselves were left undecided at the same meeting: what is agreed is the coding baseline and the requirement for a migration-friendly split, while the waveform, modulation, and numerology questions remain in study.

RAN4 sets the practical boundaries for these advances, translating features into minimum radio-frequency, resource-management, and demodulation requirements under the twin imperatives of performance and strict coexistence. The 6G spectrum paradigm is defined by scarcity: existing 5G allocations will be heavily reused, and the $\sim$7~GHz band has emerged as the leading new candidate, while 10--15~GHz remains exploratory and millimeter-wave and sub-terahertz bands are reserved for extreme localized capacity. Carrier aggregation is therefore essential, prompting a scalable requirement framework: a two-receiver configuration for legacy bands alongside a mandatory four-receiver baseline at $\sim$7~GHz, removal of legacy uplink band-combination constraints below 7.125~GHz, resource management without measurement gaps, and shared measurements across grouped devices such as a phone and a watch. Over-the-air testing has become the cornerstone for validating beamforming, AI-driven air interfaces, and non-terrestrial link dynamics.

\lesson{Spectrum access, more than waveform design, appears to be the tighter constraint}{The agreed RAN targets rest on 600~MHz of aggregated bandwidth and on reuse of the 5G site grid at around 7~GHz, and most of the RAN4 framework exists to make that aggregation workable at reasonable device cost, whereas the physical-layer decisions taken alongside were deliberately incremental. This describes where effort is currently concentrated rather than a settled conclusion: the waveform, modulation, and coding studies remain open, with constellation shaping deferred to September 2026.}

Multiple-input multiple-output (MIMO) technology completes the radio evolution. Having progressed from an advanced feature in 4G to a foundational layer in 5G, MIMO becomes an intelligent, distributed 6G enabler built on upper-mid-band spectrum, machine learning, and cell-free architectures \cite{chen23}. Massive arrays of 128--256 transceivers must overcome 6--11~dB of additional path loss at 7~GHz relative to 3.5~GHz while managing near-field propagation. At this stage, the feasibility of this approach is supported by vendor evidence rather than by 3GPP consensus, outdoor test beds show coverage parity with C-band grids alongside a 2.5-fold gain in average user spectral efficiency \cite{samsungfr3}\stag{T} and a 24~Gbps peak at 64-layer DL transmission \cite{samsung7ghz}\stag{T}, while cell-free architectures report a 20~dB coverage gain and a 3-fold capacity gain indoors \cite{onggosanusi-mimo}\stag{T}. Similarly, AI-based semantic channel-state-information (CSI) compression with a network-side decoder reportedly cuts uplink feedback overhead more than five-fold while avoiding proprietary inter-vendor encoder standardization, but this is an individual RAN1 contribution (R1-2605065)\stag{P} rather than an agreed design. Joint phase-time arrays revive millimeter-wave bands through agile sub-band analog beams, and the same distributed processing extends to sensing and satellite links. Fig.~\ref{fig:radio} maps the spectrum allocations, the June 2026 decisions, and the MIMO evolution.

\lesson{Deployability is being favored over redesign}{Cyclic-prefix orthogonal frequency division multiplexing (CP-OFDM) is retained, the SBA is evolved rather than replaced, the 5G higher-layer split may simply be inherited, and the site grid, much of the 5G spectrum, and dynamic 5G-6G sharing are all designed in. We read this as the price of deploying onto assets operators are still amortizing, which raises the burden of proof on clean-slate proposals, although the migration options themselves were left undecided in June 2026.}

\begin{figure*}[!t]
\centering
\fbox{\includegraphics[width=.985\linewidth]{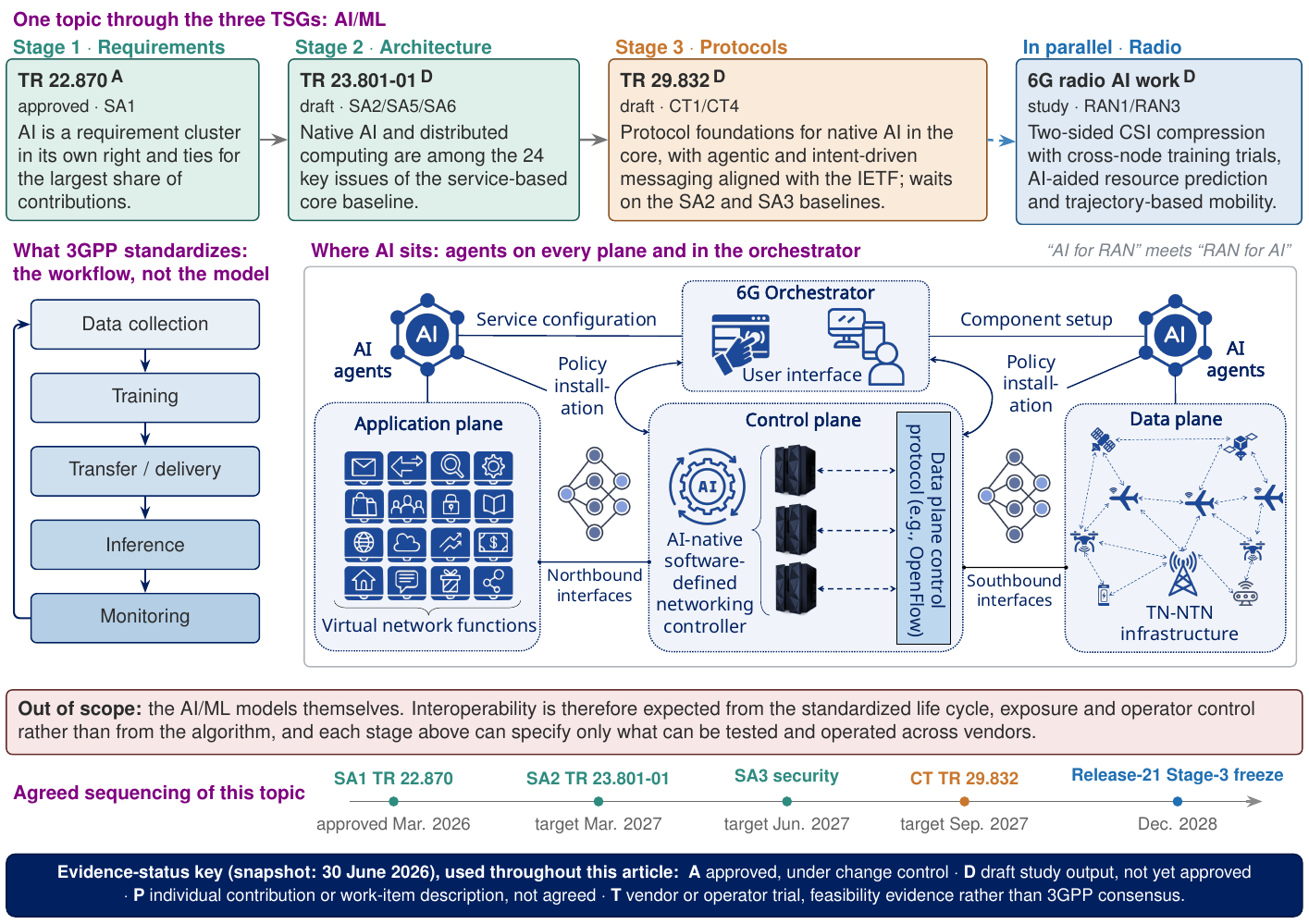}}
\caption{Native intelligence in 6G: one topic traced through the standardization machinery, and where AI sits in the system. \textit{Top:} one representative topic, AI/ML, followed from the Stage-1 service requirement in SA1, through the Stage-2 architecture in SA2 (with SA5 and SA6), to the Stage-3 protocol study in CT, with the parallel RAN air-interface work. \textit{Middle left:} the life-cycle workflow that 3GPP does specify. \textit{Middle right:} the AI-native system view, with AI agents attached to the application, control and data planes and to the orchestrator; TN-NTN denotes the combined terrestrial and non-terrestrial network infrastructure. \textit{Bottom:} the models that 3GPP does not specify, and the agreed sequencing of the four. IETF denotes the Internet Engineering Task Force. Each item carries its status at the 30 June 2026 snapshot, using the key repeated at the foot of the figure.}
\label{fig:ai}
\end{figure*}

\section{Native Intelligence: AI/ML Across the 6G System}
\label{sec:ai}
AI enters the 6G era with a substantial 5G heritage. RAN had already studied AI/ML for the NR air interface, targeting CSI feedback, beam management, and positioning, and established a functional framework for the next-generation RAN in TR~37.817\stag{A}; SA5 offers management-data-analytics services, and SA2 has evolved network data analytics since Release 15. That integration grew bottom-up, each WG applying its own understanding, and the resulting fragmentation prompted a consistency-alignment effort in TR~22.850\stag{D} to harmonize terminology, assumptions, and evaluation methods across groups \cite{chen23}.

\lesson{AI has moved from a feature to a system primitive}{In 5G, AI arrived bottom-up as isolated per-group features, which is why a consistency-alignment exercise became necessary after the fact. In Release 20 it is present at once in Stages 1, 2, and 3 and in the air interface, and it is the one topic on which nearly every group must coordinate. AI mechanisms are therefore likely to be judged on end-to-end consequences for life cycle, exposure, and governance rather than on local gain. Scope: this is an observation about where AI sits in the process, not a prediction about which techniques will be adopted.}

Crucially, the underlying AI/ML models themselves are not subject to 3GPP standardization. 
This choice exposes a structural tension between flexibility and interoperability: vendors retain freedom to optimize proprietary algorithms, yet cross-vendor validation becomes harder, particularly for two-sided models needing tight device-to-network coordination. Standardization concentrates instead on the surrounding machinery: life-cycle management, data collection, model transfer and delivery, performance monitoring, and operator control. Offline training is assumed today, with online fine-tuning the long-term target. Release-20 air-interface work advances cross-node training trials for two-sided CSI compression, AI-aided resource prediction, and trajectory-based mobility, under requirements emphasizing security, privacy, operator control, and minimal signaling overhead.

\lesson{Standardization targets the workflow, not the AI model}{AI/ML models are deliberately excluded while the machinery around them is specified in detail, so interoperability is expected from standardized workflows rather than standardized algorithms. This remains a bet rather than a demonstrated outcome: the cross-node training trials exist precisely because workflows alone do not yet guarantee that one vendor's encoder works against another's decoder.}

For 6G, the industry's ambition shifts toward a completely AI-native system, in which AI agents attach to every plane and to the orchestrator (Fig.~\ref{fig:ai}, middle right). This paradigm is captured by two complementary perspectives: ``AI for RAN'' (deploying AI to make the network smarter and more autonomous) and ``RAN for AI'' (architecting the network to serve AI workloads).

\textit{One topic, end to end.} A cross-TSG reading is easier to assert than to demonstrate, so it is worth following one topic through the whole machinery; AI/ML is the natural choice, being present in every group. In Stage~1, SA1 gives AI a requirement cluster of its own in TR~22.870\stag{A} and fixes the measurable targets, the generative-AI burst-latency and uplink figures above, that later stages must deliver and verify. Those targets enter Stage~2 as key issues in the SA2 core baseline (TR~23.801-01\stag{D}), which has to expose model and data handling as architectural services, alongside intent-driven management in SA5 and application enablement in SA6. Stage~3 turns those services into signaling: CT develops the protocol foundations for native AI in TR~29.832\stag{D}, and cannot settle them before the SA2 architecture and the SA3 security baseline are agreed, which is why its September 2027 target sits six months behind SA2. In parallel, RAN carries the air-interface half. Fig.~\ref{fig:ai} lays this path out with the status of each item and the sequencing between them, and shows why the four strands converge on one constraint: because the models stay outside 3GPP, each stage can specify only what can be tested across vendors.
Network AI agents are envisaged to interpret high-level user intent and orchestrate 3GPP services. Agentic protocols are accordingly being studied in CT so that operators can host distributed agents securely, expose network services to third-party large language models, and support natural-language operations. 
The community is candid about the open concerns: testability and interoperability, the feasibility of two-sided models, and the tendency of agentic protocols to be non-deterministic and ``chatty.'' A related debate concerns data ownership versus operator control, leaving a trade-off between centralized pipelines that are simpler to govern and federated approaches that learn better at the cost of trust. These tensions sit alongside a broader caution about ``finding problems for solutions'' amid the AI hype.
%

\begin{figure*}[!t]
\centering
\fbox{\includegraphics[width=.98\linewidth]{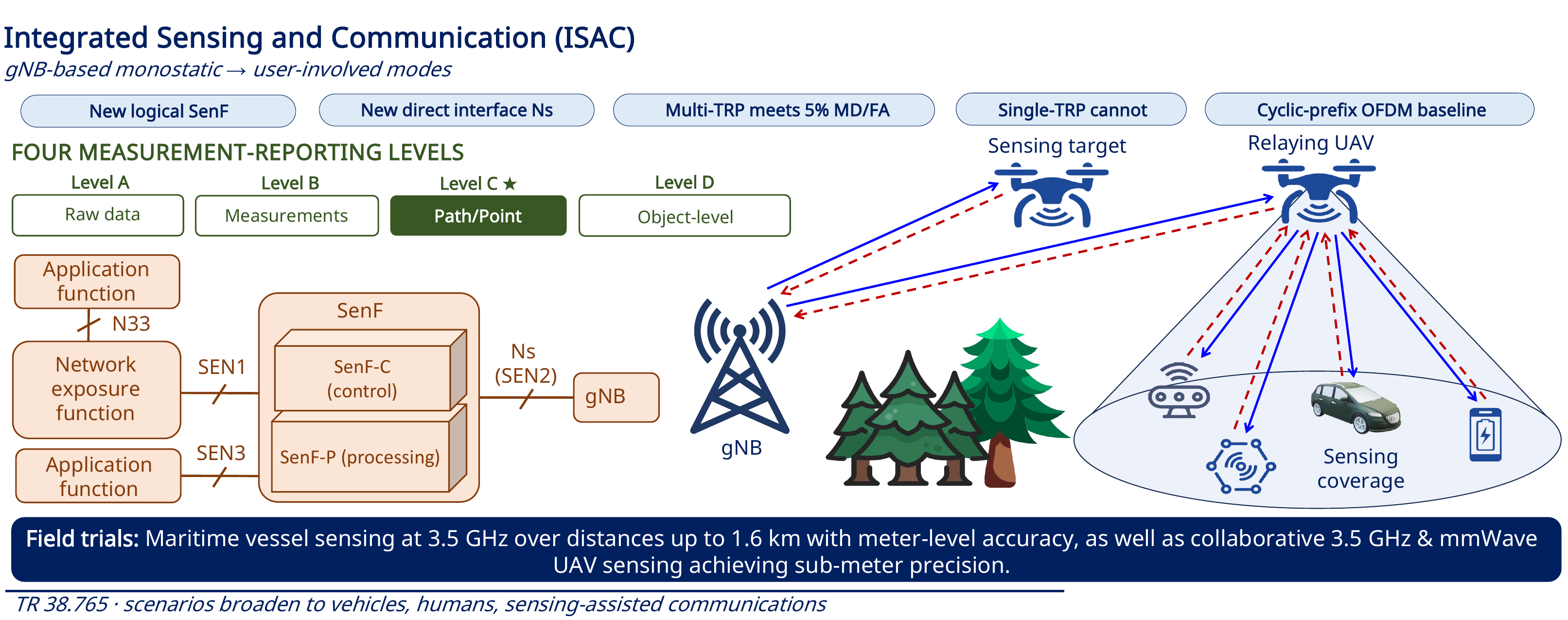}}
\caption{ISAC sensing modes, the sensing function (SenF)-based network architecture and its new Ns interface, and the four measurement-reporting levels, all drawn from the Release-20 draft study output (TR~38.765\stag{D}); the field-trial figures quoted in the lower panel are operator demonstrations\stag{T} rather than 3GPP results. In the panel, gNB denotes the next-generation base station, TRP a transmission-and-reception point, and MD/FA the missed-detection and false-alarm rates.}
\label{fig:isacntn}
\end{figure*}

\section{New 6G Capabilities: Integrated Sensing and Non-Terrestrial Networks}
\label{sec:isacntn}
ISAC is among the most tangible new capabilities of the 6G ecosystem. Following the Release-19 channel-modeling study (TR~38.901\stag{A}) in June 2025, the Release-20 phase (TR~38.765\stag{D}) evaluated base-station-based monostatic sensing for unmanned aerial vehicle (UAV) use cases with zero user-plane impact. Multi-transmission-point configurations met the strict 5\% missed-detection and false-alarm thresholds, whereas a single transmission-and-reception point fell short. Architecturally, a new sensing function (SenF), with decoupled control and processing components, connects to the base station over a new Ns interface, and of four measurement-reporting granularities evaluated the path/point level was selected for the normative phase\stag{D}. Fig.~\ref{fig:isacntn} details the sensing modes, the SenF architecture, and the measurement levels.

The Release-20 work item specifies base-station-based UAV sensing, while the broader 6G ISAC study widens the scope to user-involved topologies, vehicular and human scenarios, and a re-engineered air interface that retains CP-OFDM as its baseline without precluding chirp- or Doppler-oriented waveforms. Native sensing also raises a resource-sharing concern, since ISAC competes with communication for time-frequency resources, beamforming, and power, leaving a design trade-off between detection performance and throughput that is most acute in dense deployments. Early operator field trials point toward commercial feasibility, with maritime vessel tracking at 3.5~GHz over distances up to 1.6~km and collaborative 3.5~GHz/millimeter-wave UAV sensing; these are operator demonstrations\stag{T}, and they bear on whether the capability can be built, not on what has been agreed.

NTNs are transitioning rapidly from a supplementary 5G overlay into a native component of the 6G architecture. 
3GPP introduced foundational NR-NTN and Internet of Things (IoT)-NTN features across Releases 17 through 19 \cite{lin21space}, and the commercial satellite ecosystem is already deploying. The overarching 6G objective is a harmonized radio design supporting terrestrial networks (TNs) and NTNs ``from day one,'' a gain in capability balanced against added complexity.
Two use-case families are foreseen: direct-to-device connectivity for standard smartphones and IoT hardware below 7.125~GHz, and high-throughput links to flat-panel terminals above 10~GHz. Candidate capabilities include access without satellite navigation, autonomous positioning-navigation-timing, very-low-Earth-orbit constellations, regenerative payloads, and a dynamic three-dimensional architecture spanning geostationary and non-geostationary satellites with high-altitude platforms on optical inter-satellite links (middle panel of Fig.~\ref{fig:req}). These dimensions are driven by SA1 (TR~22.870\stag{A}), by SA2 (TR~23.801-01\stag{D}, covering interworking with earlier generations and native support for non-terrestrial links), and by RAN, and are mirrored by the ITU-R satellite track (Working Party~4B) \cite{itur}. Given heterogeneous propagation, diverse device types, and multi-orbit coordination, a genuine trade-off remains between a fully unified design and scenario-specific optimization.

\lesson{Non-terrestrial access has crossed from overlay to substrate}{Through Releases 17 to 19, satellite access was an adaptation of a design that assumed terrestrial deployment. In Release 20, harmonized support is at once an SA1 requirement\stag{A}, an SA2 core assumption\stag{D}, and a RAN design goal held from day one. Terrestrial-only system models therefore look increasingly like the special case, and the open problems have moved past feasibility to multi-orbit mobility, cross-domain synchronization, and migration from assets already in orbit. Scope: harmonization is a design goal held in the study phase, and has yet to survive normative drafting.}

\section{Core Network and Protocol Evolution}
\label{sec:ct}
The CT TSG translates Stage-2 architecture and Stage-3 security designs into the protocols, signaling frameworks, and data formats that govern device-to-core interaction, and divides its Release-20 commitment evenly between 5G-Advanced evolution and 6G studies. The pipeline opened after the CT\#111 plenary in March 2026, with 28 work-item descriptions submitted for approval\stag{P}. 
The emerging study items, all of them draft at this point\stag{D}, amount to a systematic re-examination of every core network protocol against next-generation requirements \cite{workplan,keysight_workshop}. At the signaling layer, TR~24.860 addresses evolution of the non-access-stratum protocol, while TR~29.840 revisits registration, service discovery, event exposure, and interface design. On the user plane, TR~29.841 weighs the legacy tunneling protocol against cloud-native routing alternatives such as segment routing, and a companion study targets zero-outage and self-healing core behavior. Further tracks cover network selection, capability exposure, interworking with external data networks, and a unified core data framework, while TR~29.832 covers native AI/ML in the core. Bound by inter-TSG dependencies, CT expects to conclude around September 2027, six months after the SA2 architecture and three months after the SA3 security baseline, so that Stage-3 protocols rest on stable Stage-2 foundations.

\lesson{Testability across vendors recurs as a decisive criterion}{One
constraint recurs in every domain surveyed here: models are excluded from AI
standardization while workflows are not, path/point rather than raw data was
chosen for sensing reports, harmonized TN-NTN is preferred to per-scenario
optimization, and CT accepts a six-month delay behind SA2. This is our
reading across four domains rather than an agreed 3GPP conclusion, and it rests
on a single release cycle: multi-vendor operability looks like a useful
predictor of what survives from study to specification, but it is one factor
among several.}

\begin{table*}[!t]
\renewcommand{\arraystretch}{1.2}
\caption{Cross-Cutting Lessons Supported by the June 2026 3GPP Baseline: Evidence, Observation, and Implication. Each lesson is the authors' synthesis, not a 3GPP position, and was retained only if it met the two criteria stated in Section~\ref{sec:process}. Evidence carries the status tags defined there and repeated in the footnote below.}
\label{tab:lessons}
\centering
\footnotesize
\color{black}
\begin{tabular}{|p{2.8cm}|p{6.9cm}|p{6.9cm}|}
\hline
\cellcolor{myblue}\textbf{Lesson} & \cellcolor{myblue}\textbf{Principal evidence (as of June 2026)} & \cellcolor{myblue}\textbf{Implication for the reader}\\
\hline\hline
\multicolumn{3}{|l|}{\cellcolor{myblue!45}\textit{What the system is becoming}}\\
\hline
AI has moved from a feature to a system primitive
  & Present at once in Stage 1 (tied for the largest share of SA1 contributions, TR 22.870\stag{A}), Stage 2 (SA2, SA5, SA6\stag{D}), Stage 3 (TR 29.832\stag{D}), and the air interface; TR 22.850\stag{D} corrects 5G-era fragmentation after the fact
  & AI mechanisms are judged on end-to-end consequences for life cycle, exposure, and governance, not on local gain; a technique that cannot be placed in that picture is unlikely to be adopted. Scope: a change in where AI sits in the process\\
\hline
Non-terrestrial access has crossed from overlay to substrate
  & Harmonized TN-NTN support is simultaneously an SA1 requirement (TR 22.870\stag{A}), an SA2 core assumption (TR 23.801-01\stag{D}), and a RAN design goal from day one, against overlay treatment in Releases 17--19
  & Terrestrial-only system models become the special case; the open problems are multi-orbit mobility, cross-domain synchronization, and migration from assets already in orbit. Scope: a study-phase design goal, not yet normative text\\
\hline
Connectivity is one objective among three
  & Compute-centric networking is a Stage-1 requirement cluster and SA1 splits robot service latency into communication and compute (TR 22.870\stag{A}); SA2 lists native sensing and distributed computing among its key issues (TR 23.801-01\stag{D})
  & Results stated in bits per second alone will be read against joint communication, computing, and sensing objectives; radio-only resource models are the harder case to defend. Scope: how the requirements are framed, not an agreed resource model\\
\hline
\multicolumn{3}{|l|}{\cellcolor{myblue!45}\textit{How the industry is building it}}\\
\hline
Deployability is being favored over redesign
  & CP-OFDM retained\stag{A}; SBA evolved, not replaced\stag{D}; 5G higher-layer split possibly inherited (RAN3 decision due Q3 2026\stag{P}); site grid and much of the 5G spectrum reused; dynamic 5G-6G sharing designed in
  & Clean-slate architectures carry a high burden of proof; migration paths, backward compatibility, and coexistence analysis are contributions in their own right. Scope: the migration options themselves were left undecided in June 2026\\
\hline
Spectrum access, more than waveform design, appears to be the tighter constraint
  & 36 Gbps DL and 23.4 bps/Hz underwritten by 600 MHz aggregated and site reuse at $\sim$7 GHz (TR 38.914\stag{A}); RAN4 framework built around aggregation; LDPC incremental\stag{A} and constellation shaping deferred to Sep. 2026\stag{P}
  & Proposals presuming contiguous new bandwidth are unlikely to shape Release 21; upper-mid-band coverage, aggregation, and coexistence sit closer to the critical path. Scope: waveform, modulation, and coding studies remain open\\
\hline
\multicolumn{3}{|l|}{\cellcolor{myblue!45}\textit{What decides adoption}}\\
\hline
Standardization targets the workflow, not the AI model
  & AI/ML models explicitly out of scope while the machinery around them is specified: life cycle, data collection, model transfer and delivery, monitoring, and operator control\stag{D}; cross-node trials\stag{D} and the network-side decoder proposal\stag{P} cover what workflows alone cannot
  & Interoperability is expected to arise from standardized workflows; an AI model is deployable only if its data, monitoring, and life-cycle needs fit that machinery, so report these alongside accuracy. Scope: a bet taken by the industry, not yet a demonstrated outcome\\
\hline
Testability across vendors recurs as a decisive criterion
  & AI models excluded yet workflows specified\stag{D}; path/point rather than raw-data sensing reports\stag{D}; harmonized TN-NTN over per-scenario optimization\stag{A}\stag{D}; CT staggered six months behind SA2\stag{D}
  & On the present evidence, multi-vendor operability is a useful predictor of what survives from study to specification, so evidence of it belongs in the original proposal rather than in later standardization work. Scope: our reading across four domains, from one release cycle\\
\hline
\multicolumn{3}{|p{16.9cm}|}{\footnotesize Evidence-status key (snapshot: 30 June 2026), as defined in Section~\ref{sec:process}:
\stag{A} approved by the parent group and under change control $\cdot$
\stag{D} draft output, study or normative, agreed at working-group level but not yet approved $\cdot$
\stag{P} individual contribution or work-item description, submitted but not agreed $\cdot$
\stag{T} vendor or operator trial, reported as feasibility evidence rather than as 3GPP consensus.}\\
\hline
\end{tabular}
\end{table*}

\section{Toward Trials and Commercial Launch}
\label{sec:trials}
With the study phase converging, attention is shifting to the normative push of Release 21. The freeze schedule positions the first 6G specifications to feed directly into the ITU-R IMT-2030 process, which runs parallel terrestrial (Working Party~5D) and satellite (Working Party~4B) tracks toward a 2029--2030 horizon \cite{itur}. With commercialization widely expected around 2030 \cite{lin20,chen23}, the interval between freeze and live networks is where trials, interoperability testing, and certification dominate.
Industry-led demonstration programs are already forming around it: the NTIA Mission 6G 28 initiative, for example, plans large-scale demonstrations of early 6G technologies at the 2028 Los Angeles Summer Olympics \cite{NTIA_Web}.

Several open questions will shape that interval, and the lessons collected in Table~\ref{tab:lessons} suggest which press hardest. Spectrum scarcity makes carrier aggregation and reuse of existing bands unavoidable, placing a premium on the $\sim$7~GHz upper-mid-band and on coverage techniques that match 5G's site grid. The scope of AI standardization remains a live debate: limited enough not to slow innovation, yet broad enough to guarantee testability and consistent behavior, with the role of agentic AI still being defined and two-sided models still posing inter-vendor training and validation challenges. Energy efficiency, given the RAN's share of operator energy use, is a first-class design constraint. And the harmonization of TNs and NTNs, native to 6G but demanding careful migration from in-orbit assets, will test the discipline of a single, unified design.

What unites these questions is that none is limited by the ability to design a better algorithm. Each is limited by deployability and multi-vendor operability: whether a technique runs on reused spectrum and sites, can be tested and certified across vendors, can be governed under an operator's data obligations, and can be migrated onto assets already in the field or in orbit. That is a different bar from the one much 6G research is optimized against.

\section{Concluding Remarks}
\label{sec:concl}
Seen from June 2026, 6G is no longer a vision on the horizon; it is a
coordinated body of work moving methodically from study to specification. Across the three technical specification groups, the service requirements are largely set, the architecture and security designs are taking shape, the radio interface is making its first concrete choices, native AI, integrated sensing and non-terrestrial access are being treated as native components, and the core protocols are being re-examined against 6G needs.
Release 20 carries out the studies; Release 21 will produce the first specifications and submit the IMT-2030 candidate; and the years that follow will carry the standard from the laboratory toward global trials and commercial launch around 2030.

Beyond that status, the material tells a coherent story about the character of the system being built. AI has become a primitive rather than a feature, satellite access has become a substrate rather than an overlay, and connectivity now shares the objective with computing and sensing. All three are being pursued by evolution rather than revolution, which on our reading makes spectrum and migration, rather than waveforms, the constraints that currently bite, while questions of scope tend to be settled by asking what can be tested and operated across vendors. If that reading holds, the decisive contest in 6G will be fought less over which technologies are capable than over which can be deployed at scale. It is a reading of an unfinished process, and the Release-21 work items will be its first real test.

\balance
\bibliographystyle{IEEEtran}
\bibliography{Reference}

@techreport{itur,
  author       = {{ITU-R}},
  title        = {Framework and overall objectives of the future development of
                  {IMT} for 2030 and beyond},
  institution  = {International Telecommunication Union},
  type         = {Recommendation ITU-R},
  number       = {M.2160-0},
  month        = nov,
  year         = {2023}
}

@article{lin20,
  author       = {X. Lin},
  title        = {A tale of two mobile generations: {5G-Advanced} and {6G} in
                  {3GPP} {Release~20}},
  journal      = {IEEE Commun. Standards Mag.},
  year         = {2025},
  note         = {early access, doi: 10.1109/MCOMSTD.2025.3613219}
}

@article{wang23,
  author       = {C.-X. Wang and others},
  title        = {On the road to {6G}: Visions, requirements, key technologies,
                  and testbeds},
  journal      = {IEEE Commun. Surveys Tuts.},
  volume       = {25},
  number       = {2},
  pages        = {905--974},
  year         = {2023}
}

@article{chen2026toward,
  author       = {W. Chen and others},
  title        = {Toward standardization of {6G} and {nextG}: Key technologies
                  to enable fundamental enhancements},
  journal      = {IEEE J. Sel. Areas Commun.},
  volume       = {44},
  pages        = {4333--4365},
  year         = {2026}
}

@article{lin19,
  author       = {X. Lin},
  title        = {The bridge toward {6G}: {5G-Advanced} evolution in {3GPP}
                  {Release~19}},
  journal      = {IEEE Commun. Standards Mag.},
  volume       = {9},
  number       = {1},
  pages        = {28--35},
  month        = mar,
  year         = {2025}
}

@article{chen23,
  author       = {W. Chen and others},
  title        = {{5G-Advanced} toward {6G}: Past, present, and future},
  journal      = {IEEE J. Sel. Areas Commun.},
  volume       = {41},
  number       = {6},
  pages        = {1592--1619},
  month        = jun,
  year         = {2023}
}

@article{lin21space,
  author       = {X. Lin and others},
  title        = {{5G} from space: An overview of {3GPP} non-terrestrial
                  networks},
  journal      = {IEEE Commun. Standards Mag.},
  volume       = {5},
  number       = {4},
  pages        = {147--153},
  month        = dec,
  year         = {2021}
}

@misc{samsungfr3,
  author       = {G. Xu and others},
  title        = {Upper mid-band spectrum for {6G}: Opportunities and key
                  enablers},
  howpublished = {Samsung Research Blog},
  month        = aug,
  year         = {2024},
  note         = {Accessed: Sep. 5, 2026},
  url          = {https://research.samsung.com/blog/upper-mid-band-spectrum-for-6G-opportunities-and-key-enablers}
}

@misc{samsung7ghz,
  author       = {S. Lee and others},
  title        = {Pioneering {6G} spectrum exploration: Outdoor field test for
                  extreme-{MIMO} in the {7~GHz} band},
  howpublished = {Samsung Research Blog},
  month        = mar,
  year         = {2026},
  note         = {Accessed: Sep. 5, 2026},
  url          = {https://research.samsung.com/blog/Pioneering-6G-Spectrum-Exploration-Outdoor-Field-Test-for-Extreme-MIMO-in-the-7-GHz-Band}
}

@misc{onggosanusi-mimo,
  author       = {E. Onggosanusi},
  title        = {Multiple-input multiple-output ({MIMO}) evolution: Toward
                  unlocking {6G} potential},
  howpublished = {presented at the {3GPP} {6G} workshop, Singapore, {3GPP}
                  {SGP} {WS} 09},
  month        = jun,
  year         = {2026},
  note         = {Accessed: Sep. 5, 2026},
  url          = {https://d5ln38p3754yc.cloudfront.net/content_object_shared_files/42303ffdbd0e8a42e1c6748afcdbb5aecd31ae4e/6G_3GPP_SGP_WS_09_Massive_MIMO.pdf}
}

@techreport{tr38914,
  author       = {{3GPP}},
  title        = {Study on {6G} scenarios and requirements},
  institution  = {3rd Generation Partnership Project},
  type         = {Tech. Rep.},
  number       = {TR 38.914, Release~20},
  note         = {approved at TSG RAN\#112, Jun. 2026},
  month        = jun,
  year         = {2026}
}

@misc{tsg112,
  author       = {{3GPP}},
  title        = {{3GPP} plenaries in {Singapore}},
  howpublished = {3GPP News, {TSG}\#112, Singapore, 8--12 Jun. 2026},
  month        = jun,
  year         = {2026},
  note         = {Accessed: Sep. 5, 2026},
  url          = {https://www.3gpp.org/news-events/3gpp-news/tsg112}
}

@misc{workplan,
  author       = {{3GPP}},
  title        = {{Release~20}},
  howpublished = {3GPP Work Plan},
  year         = {2026},
  note         = {Accessed: Sep. 5, 2026},
  url          = {https://www.3gpp.org/specifications-technologies/releases/release-20}
}

@misc{keysight_workshop,
  author       = {{Keysight Technologies}},
  title        = {{3GPP} {6G} standardization: From study to specification},
  howpublished = {status reports presented at the {3GPP} {6G} workshop held
                  alongside {TSG}\#112, Singapore},
  month        = jun,
  year         = {2026},
  note         = {Accessed: Sep. 5, 2026},
  url          = {https://online-events.keysight.com/series/3gpp-6g-standardization}
}

@misc{NTIA_Web,
  author       = {{NTIA}},
  title        = {Mission {6G} 28},
  year         = {2026},
  note         = {Accessed: Sep. 5, 2026},
  url          = {https://www.ntia.gov/programs-and-initiatives/mission-6g-28}
}

\end{document}